\documentstyle[12pt]{article}
\begin{document}
\title{Ambiguities in the HBT approach to determinations of interaction regions.}
\author{Kacper Zalewski\\
Institute of Nuclear Physics PAN, Krak\'ow, Poland}
\maketitle
\begin{abstract}
The necessary and sufficient condition for a quantity to be measurable by the
HBT method is given and discussed.
\end{abstract}

\section{Introduction}

The HBT method gives valuable information about the interaction regions in
multiparticle production processes, but it has important limitations. For
instance, it is well known, cf. e.g. the review \cite{LIS} and references
quoted there, that the total interaction region cannot be reconstructed. The
best one can hope for, is the reconstruction of the separate homogeneity
regions. Homogeneity region $\textbf{K}$ means the region of space, where the
particles with momentum $\textbf{K}$ are produced. The geometry of homogeneity
region $\textbf{K}$ can be described by specifying the profile function
$p(\textbf{X}|\textbf{K})$, i.e. the probability density of the production
points for the particles with momentum $\textbf{K}$. Instead of the profile
functions one could use their moments, or central moments and $\langle
\textbf{x} \rangle(\textbf{K})$, or cumulants.

In the present paper we address the question: how much can be learned about the
profile functions using the HBT method. The use of cumulants will be
particularly convenient.

\section{Simplifying assumptions}

In order to make the problem tractable and/or to avoid unessential
complications we make three simplifying assumptions.
\begin{itemize}
\item The experimental data is perfect. This means that for any $n$ the
$n$-particle momentum distribution can be measured with arbitrary precision and
fully corrected for final state interactions, resonance productions etc.
\item The profile function can be related to the emission function $S(\textbf{X},t,K)$ by the
formula
\begin{equation}\label{pverss}
  p(\textbf{X}|\textbf{K}) = \int\!\!dt\;S(\textbf{X},t,K),
\end{equation}
where the emission function can be interpreted both as a classical, time
dependent phase space distribution, then the relation is obvious,  and as a
generalized Wigner function. This assumption about the interpretation of the
emission function is made in most analyses of the HBT method. Rough estimates
suggest \cite{ZAL} that the error introduced by it is negligible for heavy ion
collisions and probably also acceptable for $e^+e^-$ annihilations. The
interpretation as generalized Wigner function is necessary to connect the
emission function, and consequently the profiles, to the measured single
particle density matrix in the momentum representation.
\item The momentum distribution for $n$ identical bosons can be expressed in
terms of an effective, $n$-independent, single particle density matrix
$\rho(\textbf{p};\textbf{p}')$ as follows \cite{KAR}

\begin{equation}\label{karczm}
  P(\textbf{p}_1,\ldots,\textbf{p}_n) = C_n\sum_P
  \prod_{j=1}^n
  \rho(\textbf{p}_j;\textbf{p}_{Pj}),
\end{equation}
where $C_n$ are normalizing constants and the summation is over all the
permutations of the momenta $\textbf{p}_1,\ldots,\textbf{p}_n$. Thus, e.g. the
single particle momentum distribution is $\rho(\textbf{p};\textbf{p})$ and for
the two-particle distribution we get the well-known formula

\begin{equation}\label{}
  P(\textbf{p}_1,\textbf{p}_2) \sim
  \rho(\textbf{p}_1;\textbf{p}_1)\rho(\textbf{p}_2;\textbf{p}_2) + |\rho(\textbf{p}_1;\textbf{p}_2)|^2.
\end{equation}
In the absence of final state interactions, resonances etc. this is a standard
assumption.

It can be made plausible as follows. $P(\textbf{p}_1,\ldots,\textbf{p}_n)$ is
the diagonal element of the $n$-particle density matrix in momentum
representation. For independent particles

\begin{equation}\label{indepp}
P(\textbf{p}_1,\ldots,\textbf{p}_n) = \prod_{j=1}^n
\rho(\textbf{p}_j;\textbf{p}_j).
\end{equation}
This corresponds to the well-known result of the theory of probabilities that
for independent events the probability of a set of events is equal to the
product of the probabilities of the single events. Bose-Einstein symmetry
implies that the $n$-particle density matrix
$\rho(\textbf{p}_1,\ldots,\textbf{p}_n;\textbf{p}'_1,\ldots, \textbf{p}'_n)$
must be invariant with respect to permutations of
$\textbf{p}_1,\ldots,\textbf{p}_n$ and with respect to permutations of
$\textbf{p}'_1,\ldots, \textbf{p}'_n$. Symmetrizing (\ref{indepp}) with respect
to the second arguments one obtains (\ref{karczm}). This symmetrization makes
the result symmetric also with respect to permutations of the first arguments,
but it spoils the normalization for $n>1$. Thus, the normalizing factors $C_n$
have to be introduced. It is seen that assumption (\ref{karczm}) can be
interpreted as the simplest symmetrized version of (\ref{indepp}).

\end{itemize}

\section{Ambiguities}

It is easily seen that the right hand side of formula (\ref{karczm}) is
invariant with respect to the substitution

\begin{equation}\label{ambigc}
  \rho(\textbf{p};\textbf{p}') \rightarrow \rho'(\textbf{p};\textbf{p}') =
  \rho(\textbf{p}';\textbf{p}).
\end{equation}
What is more interesting, it is also invariant with respect to the substitution
\cite{BIZ}

\begin{equation}\label{ambigf}
\rho(\textbf{p};\textbf{p}') \rightarrow \rho'(\textbf{p};\textbf{p}') =
e^{if(\textbf{p})}\rho(\textbf{p};\textbf{p}')e^{-if(\textbf{p}')},
\end{equation}
where $f(\textbf{p})$ is an arbitrary, real-valued function of $\textbf{p}$.
Transformations (\ref{ambigc}) and  (\ref{ambigf}) generate the full invariance
group of expression (\ref{karczm}) \cite{ZAL2}. Thus, the problem of finding
the uncertainties of the deduced profile functions reduces to the problem of
finding the implications of the ambiguities (\ref{ambigc}) and (\ref{ambigf}).

\section{Cumulants}

Let us introduce the familiar notation

\begin{equation}\label{}
  \textbf{K} = \frac{1}{2}(\textbf{p}_1 + \textbf{p}_2);\qquad \textbf{q} = \textbf{p}_1 - \textbf{p}_2.
\end{equation}
Then, using our second simplifying assumption, i.e. interpreting the emission
function as the generalized Wigner function:

\begin{equation}\label{}
  \frac{\rho(\textbf{K},\textbf{q})}{\rho(\textbf{K},\textbf{0})} = \int\!\!d^3X\;p(\textbf{X}|\textbf{K})e^{-i\textbf{qX}}.
\end{equation}
Using the terminology of the probability calculus this means that $
\frac{\rho(\textbf{K},\textbf{q})}{\rho(\textbf{K},\textbf{0})}$ at given
$\textbf{K}$ is the characteristic function of the probability distribution
(profile function) $p(\textbf{X}|\textbf{K})$.

The cumulants of profile functions will be denoted ${\cal K}(r_x,r_y,r_z)$,
where $r_x,r_y,r_z$ are non-negative integers and the argument $\textbf{K}$ is
not written explicitly. The number $r = r_x + r_y +r_z$ will be called the
order of the cumulant. The cumulant is even (odd) when $r$ is even (odd). The
cumulants are related to the characteristic function by the standard formula

\begin{equation}\label{charfu}
  \log\left(\frac{\rho(\textbf{K},\textbf{q})}{\rho(\textbf{K},\textbf{0})}\right)
  = \sum_{r_x,r_y,r_z}\frac{q_x^{r_x}\;q_y^{r_y}\;q_z^{r_z}}{r_x!\;r_y!\;r_z!}{\cal
  K}(r_x,r_y,r_z).
\end{equation}

Expanding the exponent in the integrand in formula (\ref{charfu}) one finds
that

\begin{equation}\label{}
  {\cal K}(1,0,0) = \langle x \rangle,
\end{equation}
with similar formulae for the other two order one cumulants. These are the only
cumulants which cannot be expressed in terms of the central moments of the
profile function and thus give information about the absolute positions of the
homogeneity regions. The order two and three cumulants form respectively the
covariance matrix and the skewness matrix. The fourth order cumulants are more
complicated, because they depend on both second order and fourth order central
moments. This is easily understood because, as easily checked, for a Gaussian
probability distribution all the cumulants of order higher than two vanish.

\section{Result}

Let us consider the effect of the ambiguities (\ref{ambigc}) and (\ref{ambigf})
on the deduced profile functions. Ambiguity (\ref{ambigc}) means that the data
cannot distinguish between the effective single particle density matrices
$\rho(\textbf{K},\textbf{q})$ and $\rho(\textbf{K},-\textbf{q})$. This
corresponds to the space inversion of the distribution
$p(\textbf{X}|\textbf{K})$. Since space inversion changes neither the size nor
the shape of the interaction region, this ambiguity is not very interesting.

Ambiguity (\ref{ambigf}) means that the data do not distinguish between
$\rho(\textbf{K},\textbf{q})$ and $\rho'(\textbf{K},\textbf{q})$ where:

\begin{equation}\label{}
  \log\left(\frac{\rho'(\textbf{K},\textbf{q})}{\rho'(\textbf{K},\textbf{0})}\right)
  \log\left(\frac{\rho(\textbf{K},\textbf{q})}{\rho(\textbf{K},\textbf{0})}\right)+i\left(f(\textbf{K}+\frac{1}{2}\textbf{q}) -
f(\textbf{K}-\frac{1}{2}\textbf{q})
  \right).
\end{equation}
Expanding the logarithms on both sides according to (\ref{charfu}) it is
immediately seen that the ambiguity affects only the odd cumulants. Explicitly,
the relation between the cumulants is:

\begin{equation}\label{}
  {\cal K'}(r_x,r_y,r_z) = {\cal K'}(r_x,r_y,r_z) + \left(\frac{-i}{2}\right)^{r-1}\frac{\partial^r
  f(K)}{\partial^{r_x}\;\partial^{r_y}\;\partial^{r_z}}\;\frac{1 - (-1)^r}{2}.
\end{equation}
For all real valued functions $f(\textbf{K})$ this formula gives the full
ambiguity for the cumulants and consequently also for the profile functions.
Thus, it solves our problem.

\section{Discussion}

The even cumulants ($r = r_x + r_y + r_z$  even) can be unambiguously measured
\cite{ZAL3}. In particular:

\begin{itemize}
  \item The HBT radii, which are given by the $r=2$ cumulants, can be
  unambiguously determined \cite{WIH}.
  \item The kurtosis matrix, which can be expressed by the $r=4$ and $r=2$ cumulants, can
  be unambiguously measured. The same holds for all the even central moments of
  the profile functions $p(\textbf{X}|\textbf{K})$.
  \item The distribution of $\textbf{x}_1 -\textbf{ x}_2$ for pairs of points
  in the homogeneity region, which is important for the imaging method
  \cite{BRD1}, \cite{BRD2}, \cite{BRD3}, \cite{DAP}, can be unambiguously measured,
  because it can be expressed in terms of even cumulants only \cite{ZAL3}.
\end{itemize}
The odd cumulants, which are as important for the determination of the
interaction region as the even ones, are not measurable.

In particular, the relative positions of the centers of the homogeneity region
are almost unconstrained. Since

\begin{equation}\label{xaverf}
  \langle \textbf{x} \rangle(\textbf{K}) = {\mbox{\boldmath $\nabla$}} f(\textbf{K})
\end{equation}
and $f(\textbf{K})$ is arbitrary, it is enough to assume that (for sufficiently
well-behaved functions $f(\textbf{K}))$ the rotation of the field $ \langle
\textbf{x} \rangle(\textbf{K})$ vanishes. This explains, for instance, the
behavior of the \textit{exploding source model} described in \cite{PRA}. There,
the known input-size of the interaction region is different from the size which
would be obtained by analyzing the momentum distribution under the assumptions
currently used by experimentalists. In our terminology, the function
$f(\textbf{K})$ assumed in the input is different from the corresponding
function tacitly assumed in the analysis based on the momentum distributions.
Consequently, the homogeneity regions get shifted with respect to each other
and the overall interaction region changes. In general, given a momentum
distributions for $1,2,3,\ldots$ identical bosons which correspond to a
reasonable interaction region, it is easy to construct a model where exactly
the same momentum distributions correspond to an interaction region the size of
a football.

On the other hand, the same function $f(\textbf{K})$ describes the ambiguities
for all the odd cumulants. Thus, shifting the homogeneity regions one usually
has to deform them. Consider the following example \cite{BIZ}. Assume that the
emission function is proportional to $\delta(t)$. Then the integration over
time yields the ordinary Wigner function. We start with the model where

\begin{equation}\label{}
  \rho(\textbf{p}_1,\textbf{p}_2) = \frac{1}{2\pi\Delta^2}e^{-\frac{\textbf{K}^2}{2\Delta^2} -
  \frac{1}{2}R^2\textbf{p}^2}.
\end{equation}
$\Delta^2$ and $R^2$ are positive constants. The corresponding profiles are all
Gaussian, centered at $\textbf{x}=0$. Then we deform the density matrix
according to (\ref{ambigf}) with

\begin{equation}\label{}
  f(\textbf{p}) = \frac{4}{3}\sum_{j=x,y,z}\frac{p_j^3}{a_j^3},
\end{equation}
where $a_j$ are positive constants. This model can be solved analytically and
leads to non-trivial observations. For $a_j \rightarrow \infty$ there is no
correction to the original Gaussians. When the parameters $a_j$ decrease a
number of things happens. The centers of the homogeneity regions shift so that

\begin{equation}\label{}
  \langle x_j \rangle (\textbf{K}) = -\frac{4K_j^2}{a_j^2}.
\end{equation}
This increases the overall interaction region. The main peak in
$p(x_j|\textbf{K})$ becomes steeper. To the left from it secondary maxima
develop.  Between these maxima there  are regions where the Wigner function is
negative. Negative values of the Wigner function are inconsistent with its
interpretation as a classical phase space density.  This is a well-known
problem (see e.g. the reviews \cite{TAT}, \cite{HIL}) which is usually solved
by smearing the Wigner function. A suitably smeared Wigner function can be
interpreted as a phase space density, but it is not the original Wigner
function any more. These are problems beyond the scope of the present report.

The odd cumulants, though separately unmeasurable, are strongly correlated with
each other. If the distribution $\langle \textbf{x} \rangle(\textbf{K})$ is
known, one can use equation (\ref{xaverf}) to find, up to an irrelevant
constant, function $f(\textbf{K})$ and then all the odd cumulants are
unambiguously determined. Since the even cumulants are measurable, this implies
that the profiles $p(\textbf{X}|\textbf{K})$ are also unambiguously determined.
Models based on macroscopic physics, like the hydrodynamic models, are more
likely to give correctly the distribution $\langle \textbf{x}
\rangle(\textbf{K})$ than the finer details of the shapes of the interaction
regions. The fact that getting $\langle \textbf{x} \rangle(\textbf{K})$ is
enough to determine unambiguously from experiment the profile functions may
thus be of  practical importance.

\end{document}